\documentclass[epj-spec]{svjour}
\usepackage{graphics}
\begin{document}
\title{The legacy of A.H. Wapstra and the future of the Atomic-Mass Evaluation}
\subtitle{Mass news since the last ENAM}
\author{David Lunney\inst{1}\fnmsep\thanks{\email{david.lunney@csnsm.in2p3.fr}}} 
\institute{Centre de Spectrom\'etrie Nucl\'eaire et de
Spectrom\'etrie de Masse (CSNSM)\\
IN2P3-CNRS-Universit\'e de Paris Sud, Orsay, France}
%
\abstract{This contribution pays homage to Aaldert Wapstra, the
founder of the Atomic-Mass Evaluation (AME) in its present form.
Producing an atomic-mass table requires detailed evaluation and
combination of the various decay and reaction energies as well as
data from intertial mass measurements.  Therefore, a brief summary
of all mass measurements published since the last ENAM (2004) is
given. The current status of the AME is then discussed and as well
as attempts for its continuation.
} 
\maketitle
\section{Introduction and a little history}
\label{intro}

Since the last ENAM, in 2004, a lot has happened in the field of
mass measurements. Most were good things, but there was one sad
event: In December 2006, we lost the Grand Inquisitor of the Atomic
Masses, Aaldert Hendrik Wapstra, who was 84 (see photo in Fig.~1).

\begin{figure}
\resizebox{1.0\columnwidth}{!}{
\includegraphics{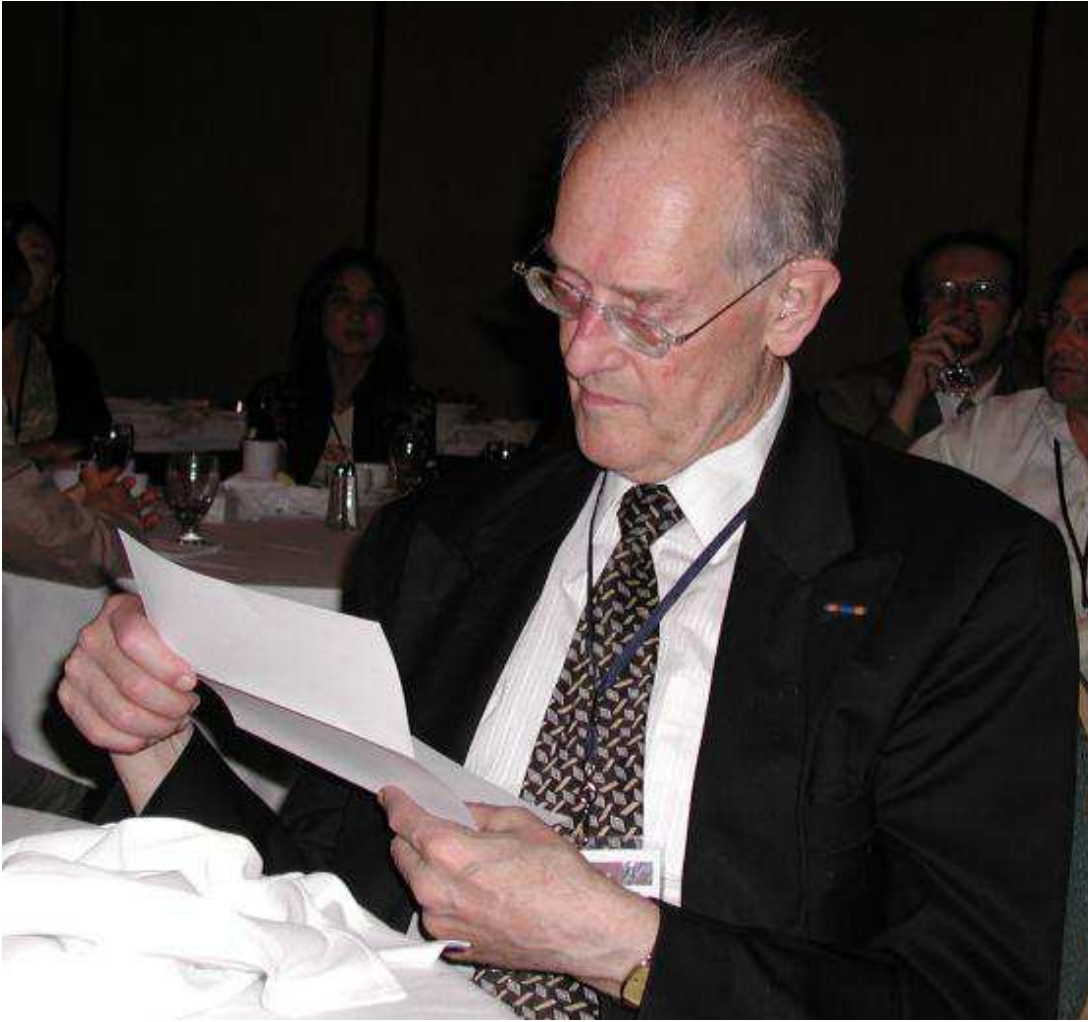} }
\caption{Photograph of Wapstra reading his SUNAMCO medal citation at
the 2004 ENAM conference.
}
\label{fig:1}       
\end{figure}

Wapstra was born in Utrecht (The Netherlands) and received his
doctorate from the University of Amsterdam in 1953.  He became a
Professor at the University of Delft in 1955 (retiring in 1987) and
was director of NIKHEF -- the premier Dutch sub-atomic physics
organization -- from 1965-1983. For details, the reader is referred
to the obituary published for him \cite{obit}. Within his 55-year
career, Wapstra created an institution in the Atomic-Mass Evaluation
-- not to be confused with a mere compilation.  The resulting ``mass
table'' is used by almost all nuclear physicists as well as
scientists in many other fields.  Wapstra opened the previous ENAM
conference so it is fitting that we started the latest ENAM by
highlighting his seminal contributions.

Masses have had a huge influence in nuclear physics and the modeling
of stellar nucleosynthesis. Shown in Fig.~2 is the gateway to the
valley of stability, the landscape of which is determined by the
nuclear binding energy. The first surveyor of this landscape was the
pioneering mass spectrometrist Francis Aston, who naturally produced
the first mass table \cite{aston}.  Aston measured masses of various
species with respect to a reference mass and established the
``whole-number rule,'' respected by all chemical species with the
exception of hydrogen. This was the first manifestation of a mass
evaluation. In fact, Aston was so obsessed by the whole-number rule
that he didn't realize he had discovered, in the small but real
deviation of hydrogen, the manifestation of Einstein's famous
$E=mc^2$ and the nuclear binding energy. Aston's contemporary, the
renown astronomer Arthur Eddington, was one of the first to realize
that the equivalence of mass and energy could be what powered the
stars. So there is a history lesson there: the link between mass
measurements and astrophysics dates from the beginning of the field!

\begin{figure}
\resizebox{1.0\columnwidth}{!}{
\includegraphics{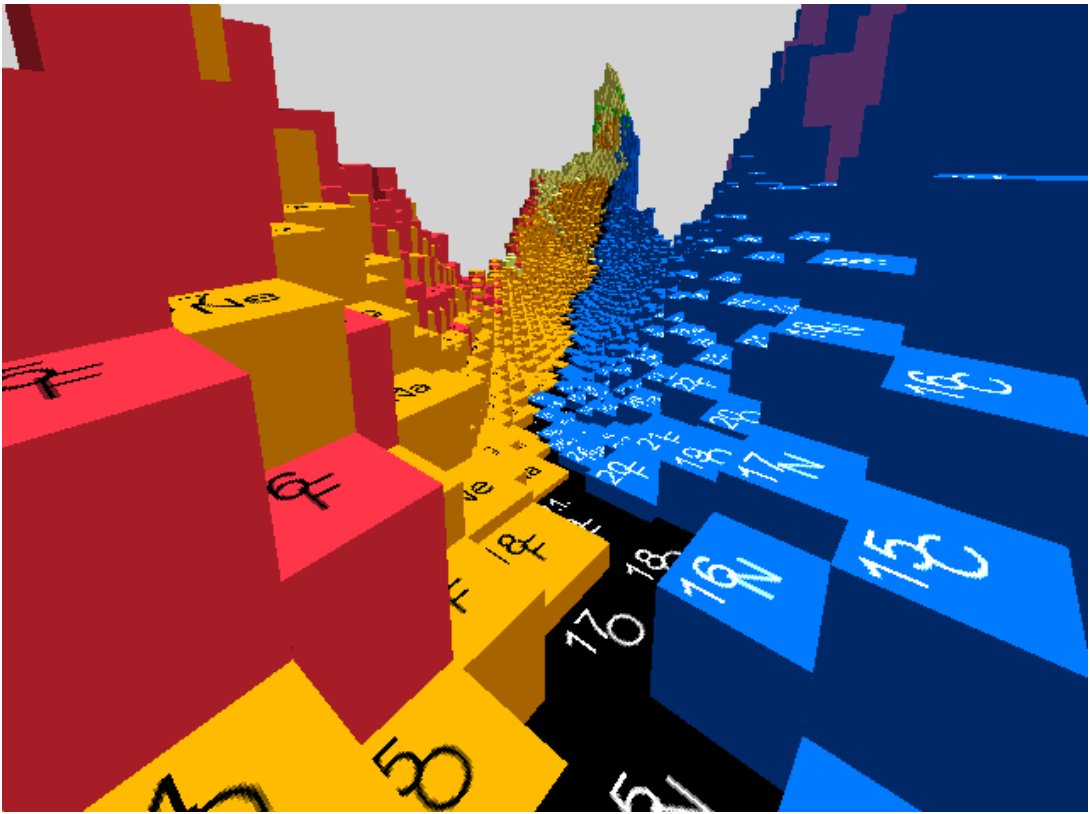} }
\caption{The gateway to the valley of stability whose landscape is
shaped by the nuclear binding energy (image by AMDC-Nucleus).}
\label{fig:2}       
\end{figure}

In the decades that followed, the field of mass measurements grew
prodigiously, hand in hand with nuclear reaction and decay studies.
Wapstra's experimental work was mostly in the field of nuclear
spectroscopy and from this large set of linked data he was first to
perform a critical evaluation of data relating to atomic masses,
transforming it all into a set of linear equations, inverting the
matix and publishing the first so-called ``mass table'' in 1960 (in
collaboration with Everling, K\"onig and Mattauch in Mainz)
\cite{ame1960}.

The publication of the 1960 table spurred a lot of activity and
already in 1961, a new evaluation was performed, this time by
K\"onig, Mattauch and Wapstra \cite{ame1961}. They showed the
differences with the previous table, many of which were larger than
their quoted uncertainties.  This revealed something we discuss a
lot today: the question of accuracy versus precision.  They also
published the links for the first time, illustrating the marked
difference between the evaluation and a straightforward compilation.

The next update was in 1964 \cite{ame1964} and the scale of the
figure showing the differences with the 1961 table was reduced by an
order of magnitude.  Further evaluations followed in 1971
\cite{ame1971}, 1977 \cite{ame1977}, and 1983 \cite{ame1983}.  This
marked the beginning of Wapstra's collaboration with Georges Audi.
This partership was to last for almost 25 years, producing four more
mass tables in the process \cite{ame1988,ame1993,ame1995,ame2003} as
well as two versions of the collection of basic ground-state
properties of known nuclides:  NUBASE \cite{nubase1997,nubase2003}.

\section{Measurement overview and comparison}
\label{sec:1}

It's interesting to look at the citations of mass evaluation: almost
a thousand (including the appendixes).  Most papers get cited only a
few times (if at all!)   I looked up several papers (using Web of
Science) and the only one close was Tanihata-san's $^{11}$Li-halo
discovery paper \cite{Tani85} with 825 citations (not including this
one).  The particle data group has more but of course that community
is much larger so I think it is safe to say that the AME is one of
the most-cited works in nuclear physics.

The strength of this so oft-cited work is all the data that goes
into it and that is, of course, provided by the many experiments -
not only mass measurement programs but also reaction and decay
measurements. Fig.~3 shows a schematic view of the different
(direct) mass measurement programs that now exist worldwide.
\begin{figure}
\resizebox{1.0\columnwidth}{!}{
\includegraphics{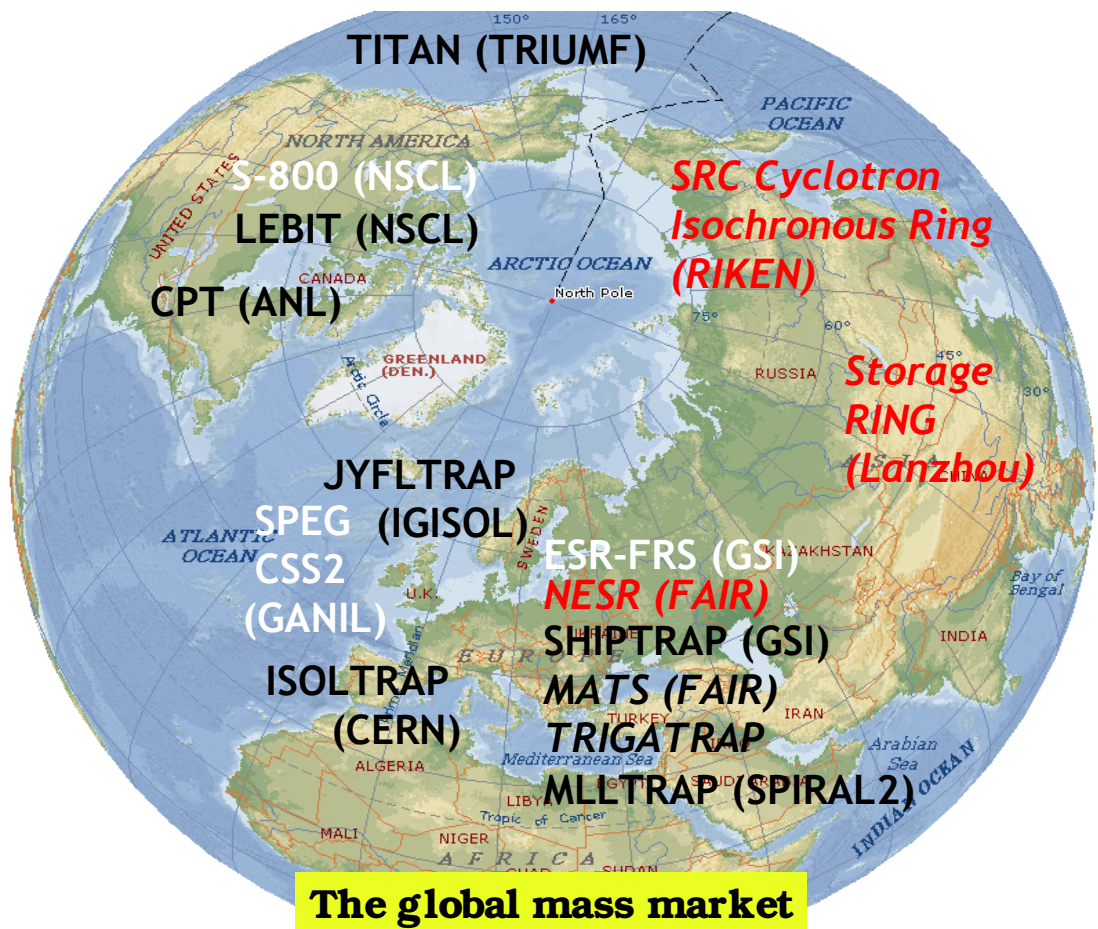} }
\caption{Illustration of how mass-measurement programs have pervaded
the world's nuclear physics installations.  (White shows
time-of-flight experiments at fragmentation facilities while black
shows Penning-trap experiments.  Others (italicized) are projects
yet to produce results.)}
\label{fig:3}       
\end{figure}

At ENAM~2004, I showed the status and the impressive progress in the
field \cite{Lunn2005}. That comparison was an extension of a review
article published less than two years earlier \cite{Lunn2003}.
(Another update was done for the 2006 Nuclei-in-the-Cosmos
conference \cite{Lunn2006}.) I thought it would be interesting to
perform the same exercise here. The progress is staggering: shown in
Fig.~4 are data published since the last ENAM -- only four short
years ago!
\begin{figure}
\resizebox{1.0\columnwidth}{!}{
\includegraphics{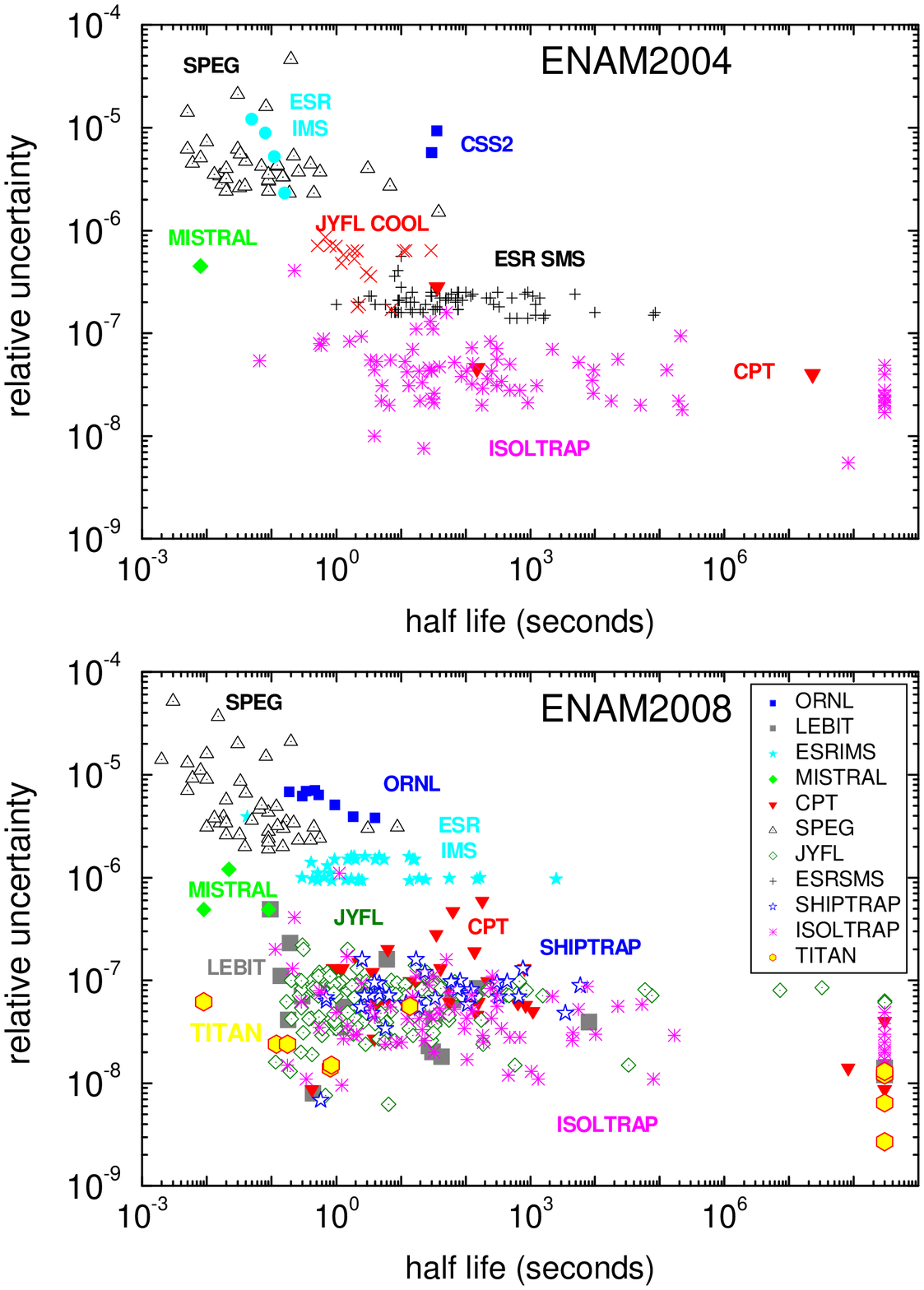} }
\caption{Comparison of mass measurements made (and published) since
that last ENAM conference in 2004.  Plotted is the mass-value
uncertainty versus half-life (source \cite{nubase2003}) for the
various direct-measurement programs (see \cite{Lunn2003},
\cite{Lunn2005}, \cite{Lunn2006} and \cite{ijms2006}).}
\label{fig:4}       
\end{figure}
There are some 330 new measurements with traps alone and those don't
include the newer ones you will have seen presented at the most
recent ENAM.  It is beyond the scope of this report to review the
different programs, especially because you should find them all
mentioned in the current special issue.  The rough break-down is as
follows: ISOLTRAP (CERN-ISOLDE):  70; CPT (Argonne): 40; SHIPTRAP
(GSI): 40; JYFLTRAP (IGISOL): 140; LEBIT (MSU): 30; TITAN
(TRIUMF-ISAC): 10.  Amongst the many highlights you will be able to
read about, I would like to single out only one:  the measurement of
the $^{11}$Li mass by TITAN \cite{smith08}.  This is, by far, the
shortest-lived nuclide ever measured with a Penning trap.  As such,
only production rate will henceforth limit the impact of these
impressive instruments.\footnote{In fitting testimony to the impact
of Penning traps on the field of mass measurements, the IUPAP
Scientific Investigator awards were awarded to the ISOLTRAP pioneers
J\"urgen Kluge and Georg Bollen (as well as acolyte Isoltrapper
Frank Herfurth in the junior category) during ENAM 2008.}

It seems obvious that we need to collect and combine it all with
existing data.  But simply putting it all in a table is not enough.
The mass evaluation maintains the links of all this data and allows
the production of a mass table as a final result. There are many
problems in nuclear structure for which this is useful, magic-number
disappearing acts being one (see O.~Sorlin, this issue). Another
important example is super-allowed beta decay, for testing
fundamental interactions (see J.C.~Hardy, this issue). Explosive
nucleosynthesis relies critically on nuclear masses (see H.~Schatz,
this issue), especially concerning the development of mass models.
On the subject of models, it is interesting to note that we also
recently lost the illustrious Carl Von Weiszaecker, author of the
pedagogical model that bears his name.  The progress of models has
been admirably chronicled by St\'ephane Goriely, Mike Pearson and
collaborators, who use the mass table as a diagnostic for exploring
their Skyrme-force parameters (see N. Chamel, this issue).

Concerning models, while you were all preparing to follow events in
Beijing, an interesting event called the Mass Olympics was organized
at the ECT* in Trento, which pitted theoreticians against each other
(though it isn't clear who the medals went to).  Some interactive
websites, allowing comparisons with different models, were also
presented (see M. Stoitsov, this issue and M. Smith, this issue).

\section{The future of the AME}
\label{sec:1}

The bulletin of the Atomic Masses Data Center (AMDC) from April
12th, 2007 contained the shocking news that the mass evaluation was
cancelled. Indeed, since that time, George Audi has not taken active
steps at pursuing the noble evaluation effort.

Combining all the spectroscopy and spectrometry data is a daunting
task.  Here are the numbers for the 2003 case:  6169 input data
which produced about 2700 masses.  But the folks at GSI invented
something even more overwhelming.  From their storage ring data come
200,000 correlations from the thousands of masses orbiting the ring
at any given time (see F. Bosch, this issue).  Because of these
correlations and the required reference masses, they would like to
combine the ESR data analysis and the evaluation into one operation.
This idea was developed within the ILIMA (Isomers, Lifetimes,
Masses) proposal for the FAIR storage rings.

Meanwhile I have undertaken a sort of mass-evaluation
apprenticeship, under the watchful eye of Georges Audi.  We have
started to collect the data and perform - I would say -
pre-evaluations and expect some additional help from GSI and the MPI
in Heidelberg. After the last ENAM conference, the Institute of
Nuclear Physics in Lanzhou (China) pledged to give the AME a
permanent home. Steps toward this laudable task will commence in
early 2009 with the target of publishing the next mass table by
2013.  The framework of the evaluation will be more open to
contributions from all Institutes and Universities. Who knows?
Perhaps one of Poland's many heros (see Fig.~5) will also help save
the evaluation!
\begin{figure}
\resizebox{1.0\columnwidth}{!}{
\includegraphics{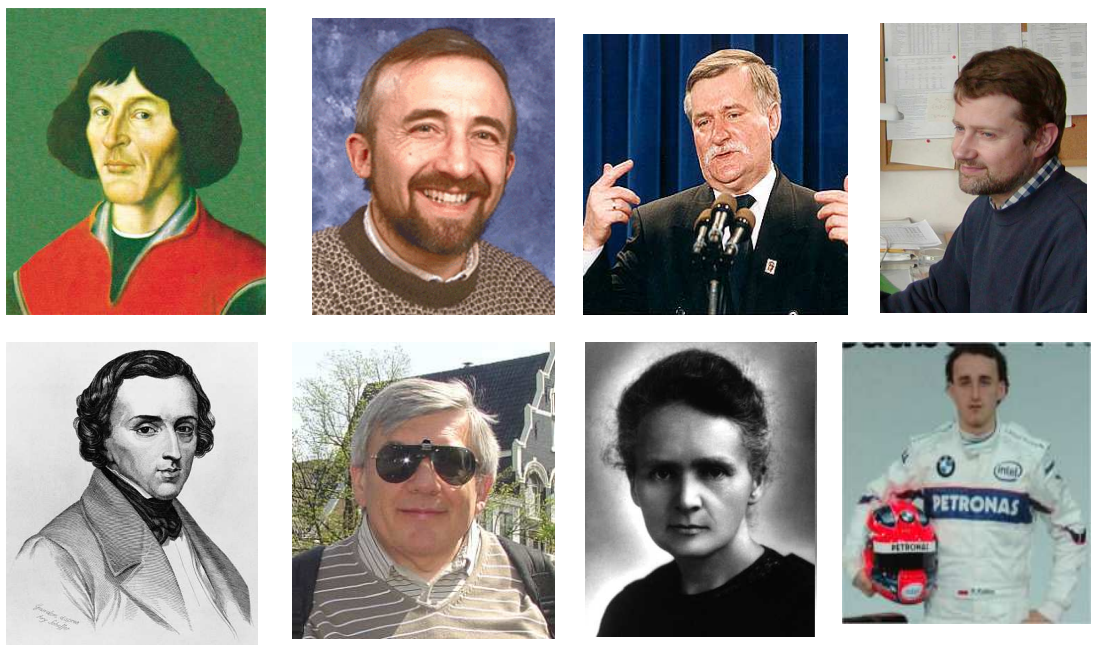} }
\caption{Some of Poland's famous personalities, some of whom may be
involved in the mass evaluation of the future.}
\label{fig:4}       
\end{figure}

\section{Conclusion}

It is more than symbolic that the first time Aaldert Wapstra didn't
make it to ENAM, the AME didn't make it to press.  We owe it to him
to make every effort so that the next ENAM will do his legacy proud.
And we will try.

%

\section{Acknowledgements}
The author thanks Georges Audi for his availability and the
organizers of ENAM~2008 for their receptiveness.

\end{document}